\documentclass[twocolumn,prl,10pt,amsmath,amssymb,nofootinbib,showpacs,superscriptaddress,floatfix]{revtex4-1}

\newcommand{\rs}{\rm\scriptscriptstyle}
\usepackage{graphicx}
\usepackage{color}
\usepackage[usenames,dvipsnames]{xcolor}
\usepackage[colorlinks=true,linkcolor=Red,citecolor=Green,linktoc=page]{hyperref}
\usepackage{multirow}
\usepackage{float}
\usepackage{flushend}
\usepackage{balance}
\usepackage[varg]{txfonts}
\usepackage{fancyhdr}

\usepackage[usenames,dvipsnames]{xcolor}
\usepackage[colorlinks=true,linkcolor=Red,citecolor=Green,linktoc=page]{hyperref}

\DeclareFontFamily{U}{rcjhbltx}{}
\DeclareFontShape{U}{rcjhbltx}{m}{n}{<->rcjhbltx}{}
\DeclareSymbolFont{hebrewletters}{U}{rcjhbltx}{m}{n}

\DeclareMathSymbol{\lamed}{\mathord}{hebrewletters}{108}

\begin{document}
\title{Superconductor-insulator transition and topological nature of the Bose metal}
\author{ M.\,C.\,Diamantini}
\affiliation{NiPS Laboratory, INFN and Dipartimento di Fisica e Geologia, University of Perugia, via A. Pascoli, I-06100 Perugia, Italy}
\author{C.\,A.\,Trugenberger}
\affiliation{SwissScientific Technologies SA, rue du Rhone 59, CH-1204 Geneva, Switzerland}
\author{I.\,Lukyanchuk}
\affiliation{University of Picardie, Laboratory of Condensed Matter Physics, Amiens, 80039, France}
\author{V.\,M.\,Vinokur}
\affiliation{Materials Science Division, Argonne National Laboratory, 9700 S. Cass Ave, Argonne, IL 60439, USA}
\affiliation{Computation Institute, University of Chicago, 5735 S. Ellis Avenue, Chicago, IL 60637, USA}

\begin{abstract}
It has long been believed that at absolute zero electrons can form only one quantum coherent state,  a superconductor. Yet, several two dimensional superconducting systems were found to harbor the superinsulating state with infinite resistance, a mirror image of superconductivity, and a metallic state often referred to as Bose metal, characterized by finite longitudinal and vanishing Hall resistances. The nature of these novel and mysterious quantum coherent states is the subject of intense study.
Here, we propose a topological gauge description of the superconductor-insulator transition (SIT) that enables us to identify the underlying mechanism of superinsulation as Polyakov's linear confinement of Cooper pairs via instantons. We find a criterion defining conditions for either a direct SIT or for the SIT via the intermediate Bose metal and demonstrate that this Bose metal phase is a Mott topological insulator
in which the Cooper pair-vortex liquid is frozen by Aharonov-Bohm interactions.
\end{abstract}

\maketitle

The dual quantum Aharonov--Bohm and Aharonov--Casher effects\,\cite{Aharonov1959,Aharonov1984} 
and a paradigm of topological phase transitions by Berezinskii-Kosterlitz-Thouless (BKT)\,\cite{Berezinskii1970,Berezinskii1971,Kosterlitz1972,Kosterlitz1973}
brought topology to prominence in condensed matter physics. The superconductor-insulator transition (SIT)  in strongly disordered superconducting films and Josephson junction arrays (JJA)\,\cite{Efetov1980,Haviland1989,Paalanen1990,Fisher1990,Fisher1990-2,fazio,Goldman2010,Diamantini1996,Doniach1998,vinokur2008superinsulator,vinokurAnnals,Fistul2008}, gripping both in a whole, hosts fascinating quantum phases emerging from the intertwined charges and vortices and their topological interactions. The nature of the SIT critical region and the phases it harbors still remain mysterious and are the subject of intense research and debate.
Here we develop a long-distance topological gauge theory of the SIT and find, in its critical vicinity, three competing quantum orders: the superconductor, the superinsulator\,\cite{Diamantini1996,Doniach1998,vinokur2008superinsulator,vinokurAnnals}, and the quantum Bose metal (QM)\,\cite{Das1999, qm}, which, as we will show, is a Mott
topological insulator (TI)\,\cite{topins2d, topins3da, topins3db}.  We find that, if quantum fluctuations are weak, the SIT occurs as a first-order direct superconductor-superinsulator transition with granular superconducting-insulating texture forming around the phase boundary. 
Strong quantum fluctuations, instead drive the SIT via the intermediate TI phase. 
The three quantum phases near the SIT epitomize the possible mechanisms for a gauge field mass without Higgs fields, with the TI and the superconductor realizing topologically massive gauge models\,\cite{jackiw} and the superinsulator implementing\,\cite{confinement} Polyakov's instanton-driven linear confinement with neutral mesons as excitations\,\cite{polyakov}. Our findings pave the way to a direct probing of gauge topological models by desktop experiments on easily accessible superconducting systems.

%%%%%%%%%%%%%%%%%%%%%%%%%%%%%%%%%%%%%%%%%%%%%%%%%%%%%%%%
\begin{figure*}[t!]
\includegraphics[width=17cm]{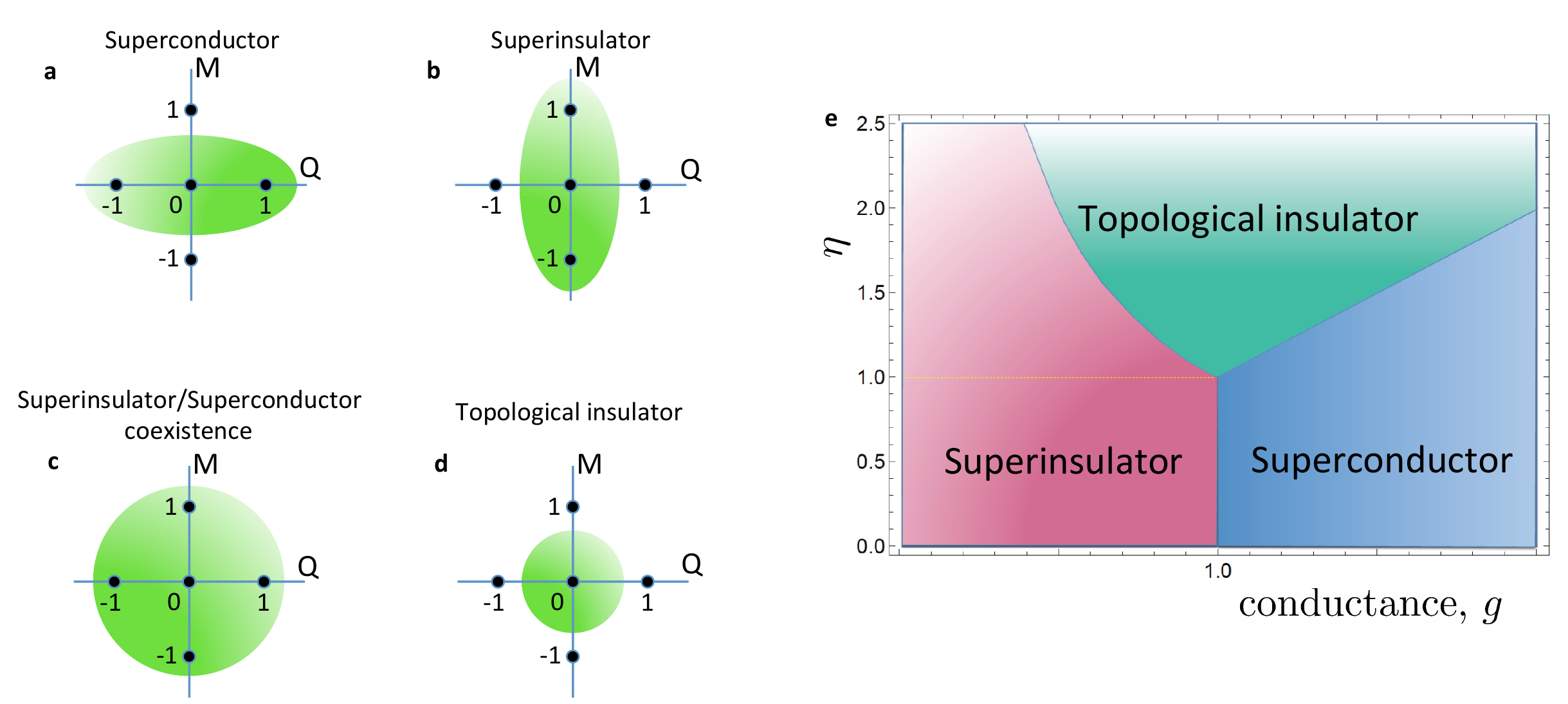}
\caption{\label{fig:Fig.1}{The configurations that minimize the string gas free energy are determined by the integer quantum numbers that fall in the interior of an ellipse with semi-axes given by the model parameters.} {(a)} Superconductor: strings with electric quantum numbers condense. {(b)} Superinsulator: Strings with magnetic quantum numbers condense. {(c)} Coexistence of long electric and magnetic strings: this is an unstable configuration near the first-order direct transition from a superinsulator to a superconductor, the strings with the quantum numbers that minimize their self-energy are the stable configurations. {(d)} Topological insulator/quantum metal: all strings are suppressed by their high self-energy. The resulting phase diagram is shown in panel (e).}
 
\end{figure*}
%%%%%%%%%%%%%%%%%%%%%%%%%%%%%%%%%%%%%%%%%%%%%%%%%%%%%%%%

In the framework of the gauge theory, the superconductor-insulator transition (SIT)\,\cite{Efetov1980,Haviland1989,Paalanen1990,Fisher1990,Fisher1990-2,fazio,Goldman2010} appears as a material realization of the field-theoretical S-duality, which, in its simplest form, expresses the invariance of Maxwell's equations under the interchange of electric and magnetic fields in the presence of magnetic monopoles\,\cite{goddardolive}. 
To gain an insight into the role of S-duality in the physics of the SIT, let us consider its simplest quantum mechanical example, a system endowed with the first-order Lagrangian\,\cite{dunne} 
$L = (\varphi \dot N - N \dot \varphi)/2 - H(N, \varphi)$, having as canonically conjugate pair of variables an amplitude $N$ and the $2\pi$-periodic phase $\varphi$. Depending on the parameters of the  Hamiltonian $H(N, \varphi)$, one finds three possible outcomes of the uncertainty principle following from the commutation relation, $[N,\varphi]=i\hbar$: (i) sharp $\varphi = x$ and plane waves $\sim e^{-ixN}$ for $H\equiv H(\varphi)$; (ii) sharp $N=y$ and plane waves $\sim e^{-iy\varphi}$ for $H\equiv H(N)$; and (iii) eigenstates with fixed quantized uncertainties $\Delta N$ and $\Delta \varphi$ in the general case. 
Going over to an infinite number of degrees of freedom, encoded in a field $\Psi=N\exp({i\varphi})$,  one finds that all three realizations of the uncertainty principle appear manifestly as distinct quantum states around the quantum tri-critical point of the SIT: a superconductor, comprising $N$-plane waves and fixed $\varphi$ (Cooper pair condensate); a superinsulator, $\varphi$-plane waves with sharp $N$ (vortex condensate)\,\cite{Diamantini1996, Doniach1998, vinokur2008superinsulator, vinokurAnnals}; and a phase harboring both frozen charge and vortex fluctuations (neither Cooper pairs nor vortices condense), which is the topological insulator\,\cite{topins2d,topins3da,topins3db}. Accordingly, the SIT is a quantum phase transition between the generic dual superconducting and superinsulating states, which can be either a direct transition or and indirect one, going through the intermediate topological insulator state.

We consider a lateral JJA, which sets the perfect stage for a gauge theory and adequately models superconducting disordered films\,\cite{Tinkham,fazio}. The properties of the JJA are controlled by the competition between the Josephson coupling energy $E_{\rs J}$ and the charging energy $E_{\rs C}=(2e^2)/2C$ of a single Josephson junction, $C$ being the junction capacitance\,\cite{Efetov1980}. At $E_{\rs J}>E_{\rs C}$, superconducting correlations win and the system is a superconductor. At $E_{\rs J}<E_{\rs C}$, the Coulomb blockade turns the system insulating. 

The Hamiltonian for a lateral JJA, modeled as a 2D grid of superconducting granules coupled by weak links is
\begin{equation}
\mathcal{\cal H}=\frac{1}{2}\sum\limits_{\mathbf x}V(C_0-C\ell^2\Delta )V+\sum\limits_{{\mathbf x},{\mathbf l}}E_{\rs J}(1-\cos(2\ell \nabla_{\mathbf l}\phi)),
\label{H1}
\end{equation}
where the sum is taken over all the points ${\mathbf x}$ of the grid, $\ell$ is the lattice spacing, ${\mathbf l}$ is the vector connecting adjacent superconducting granules, $V$ is the electric potential in the granule ${\mathbf x}$, $\nabla_{\bf l}\phi$ is the superconducting phase difference between adjacent granules, $C_0$ is the granule capacitance to the ground, and we assume $C\gg C_0$. To construct the gauge theory of the SIT, we follow\,\cite{fazio} and express the JJA partition function ${\cal Z}=\int{\cal D}V{\cal D}\varphi\exp(-{\cal H}/T)$ via an Euclidean action of a coupled Coulomb gas of charges and vortices. Then we introduce two gauge fields $a_{\mu}$ and $b_{\mu}$ which mediate these Coulomb interactions\,\cite{Diamantini1996} and, after a few transformations obtain (see Supplementary Material (SM)) the Euclidean action of a topological gauge field theory:
\begin{eqnarray}
 S=\sum_x\bigg[\frac{\ell^3}{ 8\pi^2E_{\rs J}} f_{\mu} f_{\mu} + i \frac{\ell^3}{ \pi} a_{\mu} k_{\mu \nu} b_{\nu} + \frac{\ell^3}{ 16E_c} g_{\mu} g_{\mu}+ \nonumber \\
 i\ell\sqrt{2}\left(a_{\mu} Q_{\mu} + b_{\mu} M_{\mu}\right)\bigg]\,,
\label{topdef}
\end{eqnarray}
where the sum runs over the Euclidean 3D discrete lattice, Greek indices denote the coordinate axes and repeated indices mean summation. We use natural units, $c=1$, $\hbar=1$, but we will restore the physical units when necessary. The first and the third terms in this action describe Josephson coupling and Coulomb energies in the JJA expressed through the gauge fields, $f_{\mu}=k_{\mu \nu}b_{\nu}$, 
$g_{\mu}=k_{\mu \nu} a_{\nu}$, and
$k_{\mu \nu}= S_{\mu} \epsilon_{\mu \alpha \nu} d_{\alpha}$ is the lattice Chern-Simons operator, formulated in terms of the lattice derivative $d_{\alpha}$ and shift operator $S_{\mu } f(x)= f(x +\ell \hat \mu)$ in a form guaranteeing gauge invariance (see SM). 
The second term, the so called
mixed Chern-Simons term, describes the Aharonov-Bohm coupling between
charges and vortices. Thus, the first three terms express the
doubled topologically massive gauge theory\,\cite{jackiw} for a 2D gauge
invariant massive photon with no Higgs field.  Note that, contrary to the pure Chern-Simons term used to model topological states in strong magnetic fields\,\cite{zee}, this mixed, or doubled Chern-Simons term does not violate parity and time-reversal (in Minkowski space) since it involves a vector and a pseudovector gauge field\,\cite{Diamantini1996}. 
%The topological mass $\sqrt{8E_C E_J}$ is the JJA plasma frequency. 
The last two terms, comprising integer-valued fields $Q_{\mu}$ and $M_{\mu}$, describe the topological excitations -- electric and magnetic strings\,\cite{polyakov} -- arising from the compactness of the U(1) gauge group and representing (Cooper pair) charge and vortex excitations, respectively.
Strings can be closed, in which case they are the ``world-lines" of charge-anticharge and vortex-antivortex quantum fluctuations over the ground state, or else, infinitely long, representing then the ``world-lines" of point charges and vortices. Infinitely long strings can also end on electric or magnetic (monopole) instantons, describing tunneling events\,\cite{polyakov}.

The gauge theory defined by the action (\ref{topdef}) enables a comprehensive description of the interplay of the distinct orders emerging in the critical vicinity of the SIT. Our first step is to find the free energy associated with the charge and vortex topological excitations. To that end, we integrate out the gauge fields in (\ref{topdef}) and obtain (see SM) the free energy of a topological string of length $L=N\ell$ carrying electric and magnetic quantum numbers $Q$ and $M$, respectively, assigned to every bond belonging to the string
\begin{equation}
{\cal F}=\left(\frac{1}{g}Q^2+gM^2-\frac{1}{\eta} \right) \mu\eta N\,,
\label{freeen}
\end{equation}
where we introduced the dimensionless coupling parameter $g=\sqrt{\pi^2E_{\rs J}/(2E_{\rs C})}$, with $g$$=$$g_c$=$1$ corresponding to the SIT, and the dimensionless quantity $\eta = {\pi m\ell G(m\ell)/\mu}$,  %, see the numerical plot in Fig.\,\ref{fig:Fig. 4}, 
reflecting the strength of quantum fluctuations, and $\mu\simeq\ln(5)$ is the
string entropy per bond.
Here $G(m\ell)$ is the diagonal element of the lattice Green function $G(x-y)$ representing the inverse of the operator $\ell^2 (m^2-\nabla^2)$, $m=\sqrt{8E_{\rs J}E_{\rs C}}$ is the Chern-Simons
mass, corresponding to the JJA plasma frequency, which is the inverse screening length associated with the Aharonov-Bohm interactions. 
 
If ${\cal F}$ is negative, i.e. if
\begin{equation}
(1/g) Q^2 + g M^2<1/\eta\,,
\label{ellipse}
\end{equation}
the proliferation of loops of an arbitrary size, infinitely long strings and instantons, i.e. the Bose condensation of charges and/or vortices, becomes energetically advantageous.
The condensation condition\,(\ref{ellipse}) implies that a particular condensate forms if the pair $\{Q,M\}$ on a square lattice of integer electric and magnetic charges falls within the interior of an ellipse with semi-axes  
$r_Q = (g/\eta)^{1/2}$ and $r_M=1/(g\eta)^{1/2}$. 

Figure\,\ref{fig:Fig.1}a shows the conditions for the formation of the Cooper pair condensate comprising quantum fluctuations carrying `unit' charges $\pm 2e$. In the superconducting state a charge current does not induce any voltage since the condensate of electric strings prevents the formation of vortex loops that would cause a finite resistance. In the dual, superinsulating state, see Fig.\,\ref{fig:Fig.1}b, $M=\pm 1$ and vortex-antivortex quantum fluctuations form a Bose condensate which blocks the propagation of Cooper pairs.  The conditions shown in Fig.\,\ref{fig:Fig.1}c, refer to coexisting Cooper pair and vortex condensates, $Q,M=\pm 1$. Finally, Fig.\,\ref{fig:Fig.1}d illustrates the situation where none of the condensates can form, $Q=M=0$. This is the intermediate state often referred to as a quantum (Bose) metal\,\cite{Das1999, qm}. Under the condition of exact duality and charge-hole symmetry, the Bose metal is characterized by the universal quantum resistance $R_{\rs Q}=h/(4e^2)$ at $T=0$\,\cite{Fisher1990-2}. In real materials the resistance can significantly deviate from the quantum value. 
%The phase diagram of the critical region in $g$-$T$ coordinates corresponding to $\eta<1$ (direct SIT) and $\eta>1$ (transition through the intermediate Bose metal\,\cite{Das1999, qm} which, as we show below, is a topological insulator\,\cite{topins2d, topins3da, topins3db}), are shown in Fig.\,\ref{fig:Fig.2}. 

%%%%%%%%%%%%%%%%%%%%%%%%%%%%%%%%%%%%%%%%%%%%%%%%%%%%%%%%
\begin{figure}[t!]
\includegraphics[width=9cm]{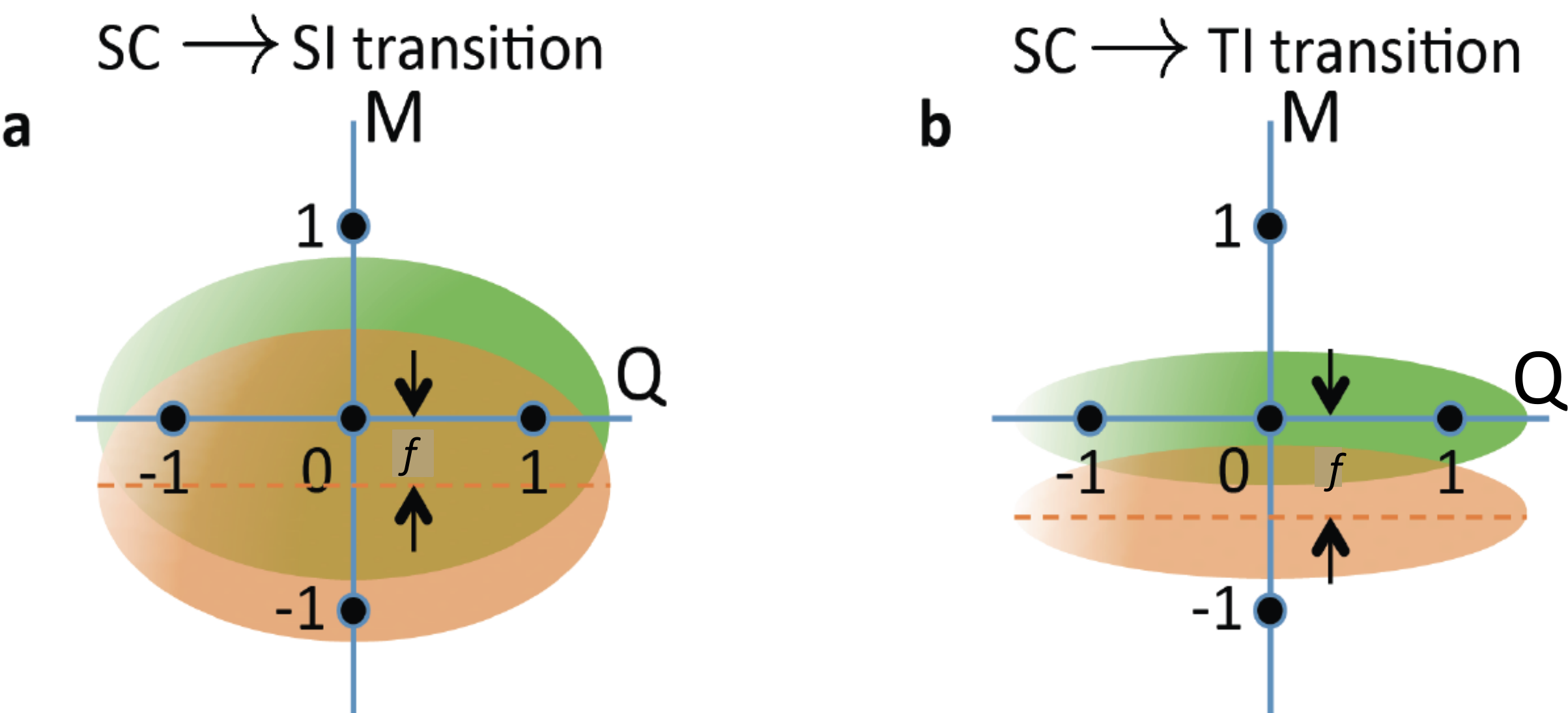}
\caption{\label{fig:Fig2} {Phase transitions induced by an external magnetic field.} The frustration parameter $0\leqslant f<1$ displaces the original ellipse along the magnetic axis: (a)\,Direct transition from a superconductor to a superinsulator for $\eta <1$. (b)\,Transition from a superconductor to a topological insulator for $\eta > 1$.}
\end{figure}
%%%%%%%%%%%%%%%%%%%%%%%%%%%%%%%%%%%%%%%%%%%%%%%%%%%%%%%%

To include the magnetic field-driven SIT in systems that are already on the brink of the SIT i.e. have $g\approx 1$, let us introduce the frustration factor $f=B/B_{\rs \Phi}$, where $B_{\rs \Phi}$ is the magnetic field corresponding to one flux quantum $\Phi_0=\pi\hbar c/e$ piercing one plaquette of the JJA. An external magnetic field corresponds to a special case of condensed magnetic strings with non-integer quantum number and results thus in a shift of the magnetic quantum number in the free energy and, accordingly in the condensation condition\,(\ref{ellipse}): $M\to M+f$. This shift along the magnetic axis modifies the condensation conditions, see Fig.\,\ref{fig:Fig2}. Let us assume $g=1+\epsilon$, where $\epsilon\ll 1$, so that the system is in a superconducting state but close to the SIT. Then for a direct SIT at $\eta<1$ one finds $f_{\rm c} = (1/2)(g^2-1)\approx\epsilon$.
At $\eta>1$, but still close to the tri-critical point, we set $g=\eta+\epsilon$ and find that the superconductor transforms into a topological insulator at $f_{c}=\sqrt{\epsilon}/\eta^{3/2}=(g-\eta)^{1/2}/\eta^{3/2}$.

%%%%%%%%%%%%%%%%%%%%%%%%%%%%%%%%%%%%%%%%%%%%%%%%%%%%%%%%
\begin{figure}[t!]
\includegraphics[width=8cm]{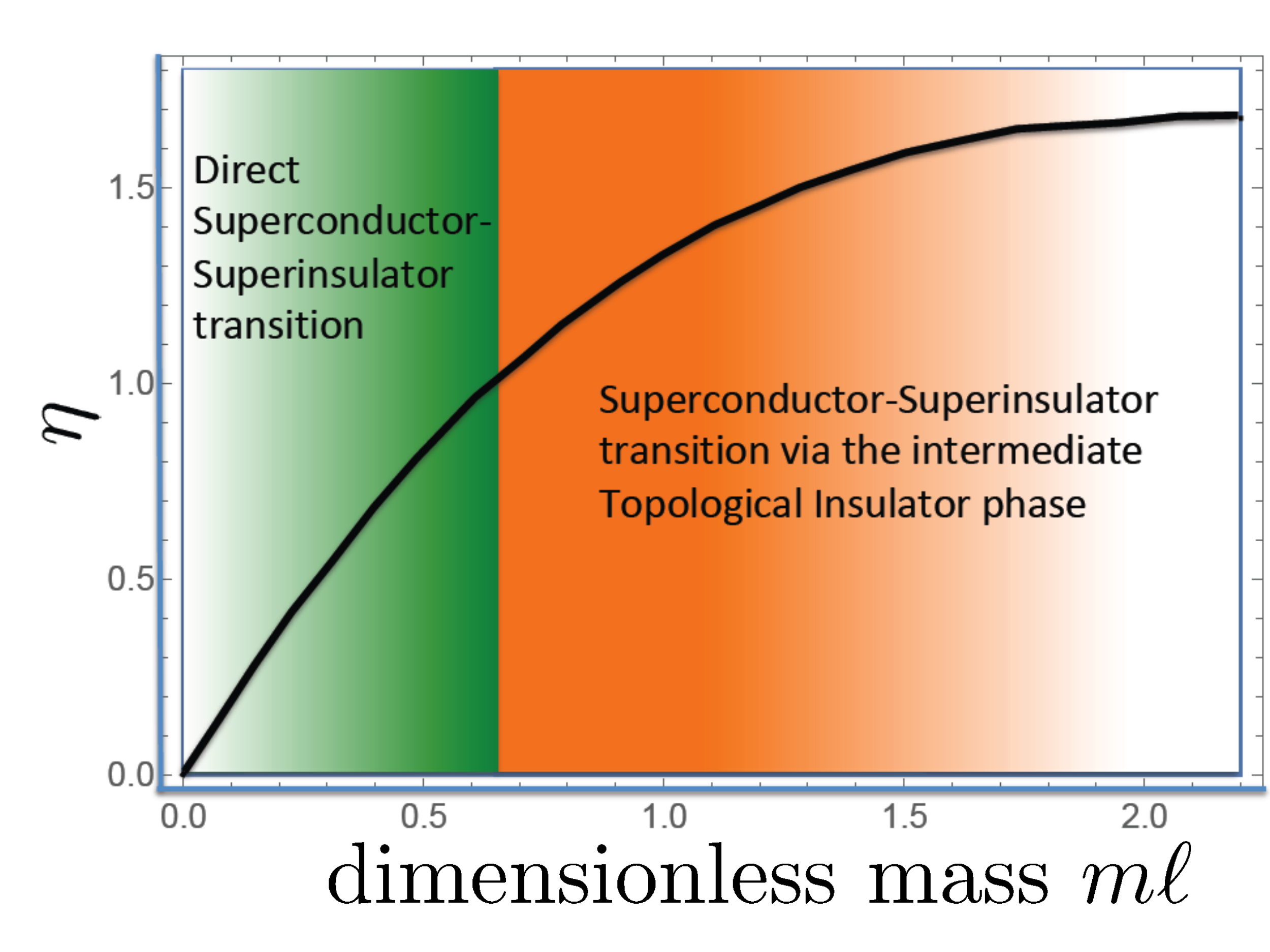}
\caption{\label{fig:Fig.3} {Representative plot of $\eta$ as a function of the parameter $m\ell$ for $\mu = 0.25 \ {\rm ln}(5)$.} The critical value of $m\ell$ that separates the domains of weak and strong quantum fluctuations is $(m\ell)_{\mathrm c}=0.65$. At $m\ell>0.65$, quantum fluctuations are strong enough so that $\eta>1$ and the SIT at $T=0$ occurs via the intermediate topological insulator phase.}
\end{figure}
%%%%%%%%%%%%%%%%%%%%%%%%%%%%%%%%%%%%%%%%%%%%%%%%%%%%%%%%

We now extend the gauge theory of the SIT and the analysis of the corresponding emergent phases onto disordered superconducting continuous media. To do so, we transcribe the above results obtained in terms of the JJA parameters $E_{\rs C}$ and $E_{\rs J}$ into the language of material characteristics of superconducting films. The role of the tuning parameter driving the film across the SIT at zero magnetic field is taken by the resistance per square, $R_{\square}$ (or, equivalently, by the dimensionless conductance $g=4R_{\rs Q}/R_{\square}$). The disorder-driven SIT in films is expected to occur at $g=g_c=1$ which corresponds to the condition ${\sqrt{2E_{\rs C}/\pi^2E_{\rs J}}}= 1$. A further connection between the material characteristics of the superconducting films and those of JJA is established by the relation\,\cite{Tinkham} 
$\lambda_{\perp}=c\Phi_0/(8\pi^2I_c)$, where $\lambda_{\perp}=\lambda_{\rs L}^2/d$ is the Pearl screening length, $\lambda_{\rs L}$ is the bulk London penetration depth, $d$ is the film thickness, and $I_c\equiv(2eE_{\rs J}/\hbar)$ is the critical current of a single Josephson junction. Then, our dimensionless parameter $m\ell$ acquires the form 
\begin{equation}
m\ell=\frac{1}{8\alpha} {1\over \kappa} \ ,
\label{universal}
\end{equation}
where $\alpha=e^2/(\hbar c)$ is the fine structure constant, $\ell$ now plays the role of the characteristic microscopic cutoff length of order of the superconducting coherence length $\xi$ and thus $\kappa = \lambda_{\perp}/\ell$ is the (2D) Landau parameter of the superconducting film. Finally we have to generalize (\ref{topdef}) to its non-relativistic version containing material parameters such a magnetic permeability $\mu_P$ and a dielectric constant $\epsilon_P$ with light velocity in the medium given by $v=1/\sqrt{\mu_P \epsilon_P}$ (see SM). We note here that JJA correspond formally to the limit $\mu_P \to \infty$ with $\epsilon_P = 1$ (see SM).  For the superconducting films of interest in this paper, on the contrary, only the dielectric constant is a relevant parameter while we can safely set $\mu_P=1$. 
In this case the parameter $\eta$ decomposes into three factors (see SM)
\begin{equation}
\eta = {1\over \alpha} {1\over \kappa} {\pi v^2 \tilde G(ml) \over 8\mu} \ .
\label{etaform}
\end{equation}
The first, universal factor indicates that the parameter $\eta$ describes the strength of quantum fluctuations: an intermediate quantum metal/topological insulator phase occurs when these are strong enough, i.e. $\eta >1$. The second and third factors embody all relevant material parameters, namely the Landau parameter of the superconducting substrate and the dielectric constant. This latter appears only in the third term, where all non-relativistic effects are bundled. 
In disordered superconductors, where the coherence length $\xi$ exceeds the mean free path, $\lambda_{\rs L}^2=[4\pi n_se^2/(mc^2)](\tau\Delta/\hbar)$, where $n_s$ is the density of superconducting electrons, which at $T=0$ is equal to the total electron density $n_e$ and $\tau$ is the transport scattering time. Accordingly, $m\ell \simeq(\alpha/4)(\ell dn_e^{2/3})(\tau\Delta/\hbar)$. If we associate the ultraviolet cutoff $\ell$ with the minimal superconducting scale $\xi$, the dependence $m\ell \sim n_e^{2/3}$ reflects correctly the trend of crossing over from the direct SIT to SMIT upon an increase of the electron density in films.

Taking $\mu = {\rm ln}(5)$ and an example for which $r=v^2\tilde G(ml)/G(ml) = 4$, see Fig.\,\ref{fig:Fig.3},  the magnitude of the parameter $(m\ell)_c$ where all three phases meet, i.e. its magnitude corresponding to the tri-critical point, is $(m\ell)_c\approx 0.65$. However, when describing real superconducting films one has to take into account that $\varepsilon$ and thus also $v$ and the above parameter $r$ are not fixed but, rather depend on the proximity of the system to the SIT and that $\varepsilon$ diverges as $g\to 1$\,\cite{vinokurAnnals}. Since $\eta$ decreases with increasing $\varepsilon$, the scenario of the SIT, direct vs. via the intermediate quantum metal phase, depends on the magnitude of $\eta(\varepsilon_\mathrm{max})$ where $\varepsilon_\mathrm{max}$ is the maximal dielectric constant achieved at the value of $g=1$, where the correlation length associated with the SIT compares to the lateral dimension of the film. Taking an estimate of $\varepsilon_\mathrm{max}\simeq 10^4$ as a characteristic value for the NbTiN film, where the divergent $\varepsilon$ was observed on approach to the magnetic field-driven SIT\,\cite{Mironov:2017}, and using a corresponding estimate of $m\ell \approx 0.26$ for this material one obtains $\eta < 1$ and hence one expects this NbTiN film to exhibit a direct SIT, as indeed is observed experimentally. Analogously, one can observe that near the SIT $\eta<1$ for TiN, for which $m\ell \approx 2.63$, which thus also follows the direct SIT scenario, as observed. Instead, for NbSi, one does not expect a $\varepsilon$ divergence and, correspondingly, the estimate $m\ell \approx 21.6$ gives $\eta > 1$ confirming a transition via an intermediate quantum (Bose) metal phase.

To unravel the nature of the intermediate quantum Bose 
metal, whose mysterious nature is the subject of intense scrutiny\,\cite{Kapitulnik:2016,Phillips:2002,Phillips:2003,Phillips:2017,Dalmonte:2017,Kapitulnik_Rev2017,Kapitulnik2017}, 
we return to the gauge action\,(\ref{topdef}).  When neither Cooper pairs nor vortices condense, they experience the mutual statistical repulsion\,\cite{wilczek} 
due to the topological Aharonov-Bohm/Aharonov-Casher effects and get thus frozen in a ``doubled version" of an incompressible quantum fluid. As a consequence, the bulk dynamics is completely suppressed and  the flow of charges and vortices is supported only by edge excitations. To see this, one has to calculate the induced charge current
$j^{\mu}_{\rm ind} =(1/ \ell^3)[\delta S^{\rm eff}( A_{\mu})/\delta A_{\mu}]$, 
where $A_{\mu}$ 
is the electromagnetic potential  
coupled to the probe electric charge.
In the phase where topological excitations do not condense, the kinetic terms in the action (\ref{topdef}) become irrelevant and the whole action is dominated by the topological doubled Chern-Simons term\,\cite{moore}, leading to an effective action in which the topological Chern-Simons mass suppresses all bulk conductances: 
\begin{equation}
S^{\rm eff} \left( A_{\mu}\right) = \sum_x {\ell^4 e^2 g \eta \mu\over  \pi^2} A_{\mu } \left( -\delta_{\mu \nu} \nabla^2 + d_{\mu} \hat d_{\nu} \right) A_{\nu} \ .
\label{topinseff}
\end{equation}
This is the effective action of a bulk insulator. The functional derivative with respect to the field $A_{\mu}$ yields the current induced by external electromagnetic fields $F_{\mu \nu}=\partial_{\mu}A_{\nu}-\partial_{\nu}A_{\mu}$ as 
\begin{equation}
j^{\mu}_{\rm ind} = \ell {2 e^2 g\eta \mu\over \pi^2} \partial_{\nu} F^{\mu \nu}\,,
\label{indcur}
\end{equation}
showing that only \textit{variations} of external electric and magnetic field, but not homogenous field themselves, can cause an induced current, which is a typical electromagnetic response of an insulator. Therefore,  the bulk conductances, both the longitudinal and the Hall ones, vanish.

%%%%%%%%%%%%%%%%%%%%%%%%%%%%%%%%%%%%%%%%%%%%%%%%%%%%%%%%
\begin{figure*}
\includegraphics[width=16cm]{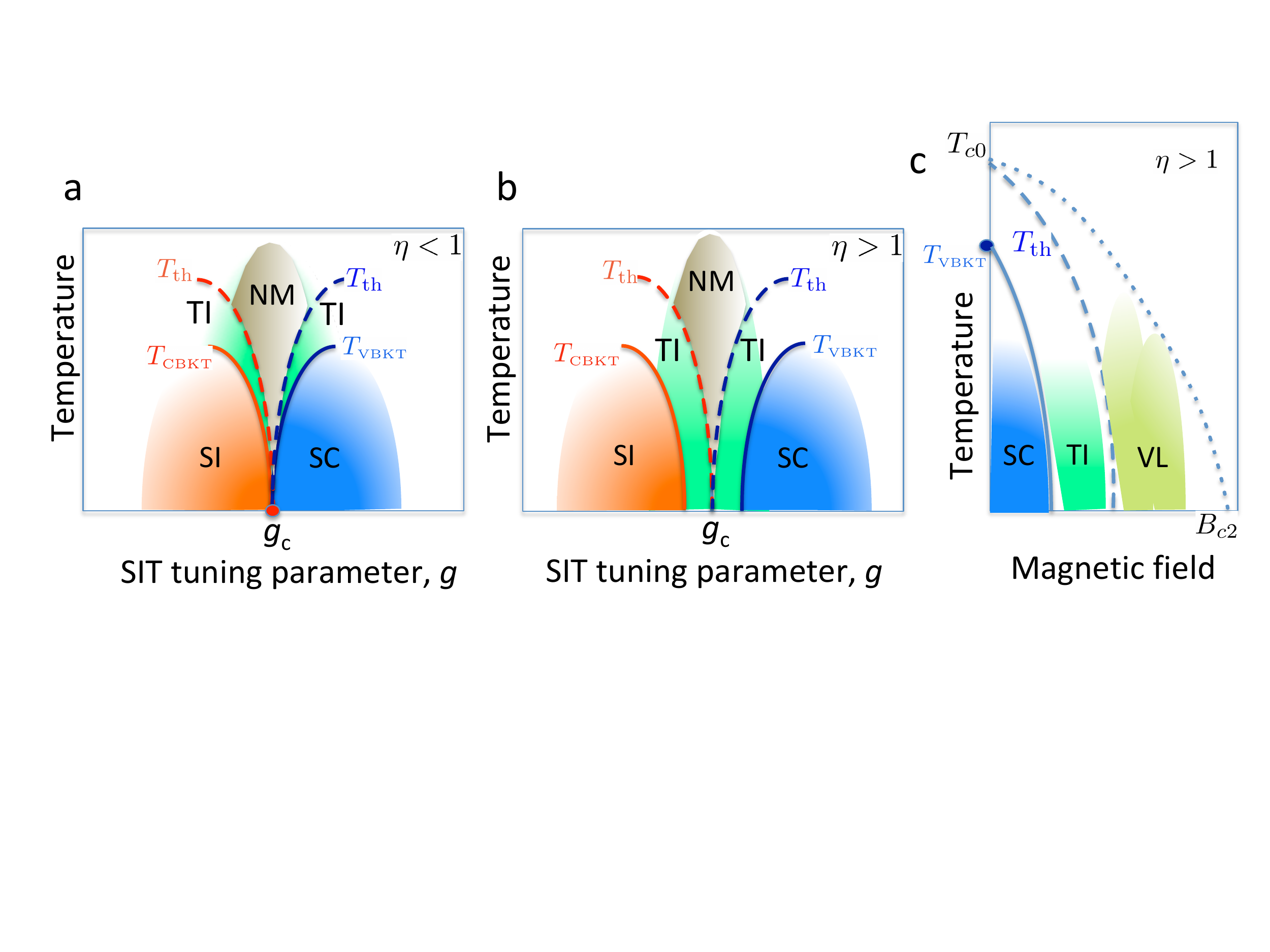}
\caption{\label{fig:Fig.4} {Phase diagram of the vicinity of the SIT.} Tuning the parameter $g=\sqrt{\pi^2 E_{\rs J}/2E_{\rs C}}$, one drives the system across the SIT. The quantity $\eta$ characterizes the strength of quantum fluctuations. Panels (a) and (b) show the SIT in zero magnetic field. (a)\,Weak quantum fluctuations, $\eta<1$, where the system experiences the direct SIT at $g=g_c$ and $T=0$. (b)\,If quantum fluctuations are strong, $\eta > 1$, an intermediate quantum state, Bose metal, which is demonstrated to be a topological insulator, opens up between the superconductor and the superinsulator at $T=0$. The solid lines separating phases are deconfinement lines coinciding with the BKT transitions lines. The dashed lines are thawing lines where the frozen Chern-Simons topological insulator melts. At $B=0$ thawing lines coincide with the superconducting, $T_c$ and insulating, $T_{I}$ transition temperatures, respectively. At finite magnetic field Chern-Simons phase melts into either conventional insulator or vortex liquid (weak metal). (c)\,Sketch of the phase diagram in the $T$--$B$ coordinates at the superconducting side and $\eta>1$.  }
\end{figure*}
%%%%%%%%%%%%%%%%%%%%%%%%%%%%%%%%%%%%%%%%%%%%%%%%%%%%%%%%

This is not the whole story though. The Chern-Simons effective action is not invariant under gauge transformations $a_i=\partial_i \lambda$ and $b_i=\partial_i \chi$ at the edges. Two chiral bosons\,\cite{flore} $\lambda = \xi + \eta$ and $\chi = \xi -\eta$ have to be introduced to restore the full gauge invariance (see SM), exactly as it is done in the quantum Hall effect framework\,\cite{qhe}. The full gauge invariance is restored by adding the edge action (we use continuum notation for simplicity's sake) 
\begin{equation}
S_{\rm edge} = {1 \over \pi} \int d^2 x \ \left( \partial_0 \xi \partial_s \xi  - \partial_0 \eta \partial_s \eta \right) + 
2e\int d^2 x \  A_0  \left( {\sqrt{2}\over 2\pi}\partial_s \chi \right) \ ,
\label{edgeac}
\end{equation}
where we have included the electromagnetic coupling of the edge charge density $\rho= (\sqrt{2}/ 2\pi)\partial_s \chi $
in the $A_s=0$ gauge. 
As in the case of the quantum Hall effect, the non-universal dynamics of the edge modes is generated by boundary effects \cite{qhe}, which result in the Hamiltonian
\begin{equation}
H = {1 \over \pi} \int ds \left[ -v \left( \partial_s \xi \right)^2 -v \left( \partial_s \eta \right)^2\right] \ ,
\label{four}
\end{equation}
where $v$ is the velocity of propagation of the edge modes along the boundary. Upon adding this term, the total edge action becomes
\begin{eqnarray}
S_{\rm edge} = {1 \over \pi} \int d^2 x \ \left[ \left(\partial_0-v\partial_s\right) \xi \partial_s \xi  - \left(\partial_0+v \partial_s \right)\eta \partial_s \eta \right]\nonumber \\ + 2e \int d^2 x \  A_0  \left( {\sqrt{2}\over 2\pi}\partial_s \chi \right) \ . 
\label{edgeacfull}
\end{eqnarray}
The equation of motion generated by this action is 
\begin{equation}
v \partial_s \rho= {2e \over 2\pi} E = {2e \over 2\pi} \partial_s A_0\ ,
\label{eight}
\end{equation}
i.e. the charge conduction with the quantum resistance $R_{\rm Q}=h/(4e^2)$. Thus, while all bulk conductances are suppressed by the large Chern-Simons mass, there remains the ballistic edge conductance with the quantum resistance,  
which is characteristic of a topological insulator\,\cite{topins2d, moore}, in this case a Mott topological insulator. The edge modes in this topological insulator carry the charge $2e$ since the elemental carriers mediating the conductivity are Cooper pairs. This is reminiscent of recent observations of doubly charged egde excitations in the integer quantum Hall effect regime\,\cite{iqh}. Then, a straightforward calculation (see SM) yields the TI sheet resistance $R_{\square}=h/(4e^2)\equiv R_{\rs Q}$, which corresponds to perfect duality, where exactly one fluxon for each Cooper pair traverses the system. This coincides with the result by M.\,P.\,A.\, Fisher\,\cite{Fisher1990} obtained by an elegant qualitative consideration.  
%This boundary field theory induced by a topological Chern-Simons effective field theory in the bulk is the simplest example of the holographic principle as applied to condensed matter systems\,\cite{zaanenbook}. 
In experiments, the resistance can significantly differ from $R_{\rs Q}$ signaling deviations from strict duality\,\cite{Kapitulnik_Rev2017}.
Remarkably, our prediction that in the Bose metal the Hall resistance should disappear is in full accord with the recent observation by\,\cite{Kapitulnik2017}.

We further expect that, at the temperature $T=T_{\mathrm{th}}$ associated with the topological mass gap, thermal fluctuations become strong enough so that the frozen state of intertwined, non-condensed vortices and Cooper pairs would thaw and the TI transforms into an ordinary bulk insulator or into a vortex liquid (bad metal). Taking into account non-relativistic effects due to material parameters, the energy gap associated with the Chern-Simons mass in the critical vicinity of the SIT is (see SM)
$m/\sqrt{\epsilon} \propto(g-g_c)^{\nu/2}$, where $\nu$ is the scaling exponent describing the divergence of the dielectric constant of the film on approach to the SIT\,\cite{vinokurAnnals}. Then, by the the duality principle\,\cite{Fisher1990,Fisher1990-2}, one would expect the Chern-Simons melting temperature to scale as $T_{\mathrm{th}}\propto |g-g_c|^{\nu/2}$, as shown in Fig.\,\ref{fig:Fig.4} by dashed lines.

A qualitative description of the transformation of the superconducting/superinsulating phases into the TI and further into a bulk insulator/vortex liquid is achieved by including the magnetic field driving the transition into our consideration. To that end, let us first introduce the dimensionless temperature $\tau=T\ell/(\hbar v_c)$, with $v_c$ the light velocity in the material. As derived in a previous section, the critical field for the superconductor-to-TI transition is $f_c=\sqrt{\epsilon}/\eta^{3/2}=(g-\eta)^{1/2}/\eta^{3/2}$. The important point is then that, at finite temperatures, the parameter $\eta$ scales as $\eta(T) =\eta_0 S(T)$, with a scale factor $S(T) >1$ and $\eta_0=\eta(T=0)$, as derived in \cite{confinement}. Using this scaling we obtain 
the equation that defines the SC$\leftrightarrow$TI transition temperature $\tau_{\rs TI}$ as a function of the applied magnetic field,
\begin{equation}
f(\tau_{\rs TI})=\frac{\sqrt{\eta_0(1-S(\tau_{\rs TI}))+\epsilon_0}}{\eta_0^{3/2}S^{3/2}(\tau_{\rs TI})} \ ,
\end{equation}
where $\epsilon_0=g-\eta_0$. The critical temperature for the SC$\leftrightarrow$TI transition at $f=0$ is then defined by the equation
\begin{equation}
S(\tau_{\rs TI})=1+\epsilon_0/\eta_0 \ .
\end{equation}
The dimensionless thawing temperature, instead, is determined by the topological mass splitting caused by the external magnetic field (see SM),
\begin{equation}
\tau_{\mathrm{th}}=\frac{m\ell}{\hbar v_c}(1-f/2) \ ,
\end{equation}
and the intermediate phase appears at zero magnetic field if $m\ell/(\hbar v_c)>\tau_{\rs TI}$. The sketch of the resulting SC$\leftrightarrow$TI transition phase diagram in $T$-$B$ coordinates is shown in Fig.\,\ref{fig:Fig.4}c and reproduces the corresponding experimental phase diagram of\,\cite{Kapitulnik2017}. 

%%%%%%%%%%%%%%%%%%%%%%%%%%%%%%%%%%%%%%%%%%%%%%%%%%%%%%%%
\begin{figure}[t!]
\includegraphics[width=8cm]{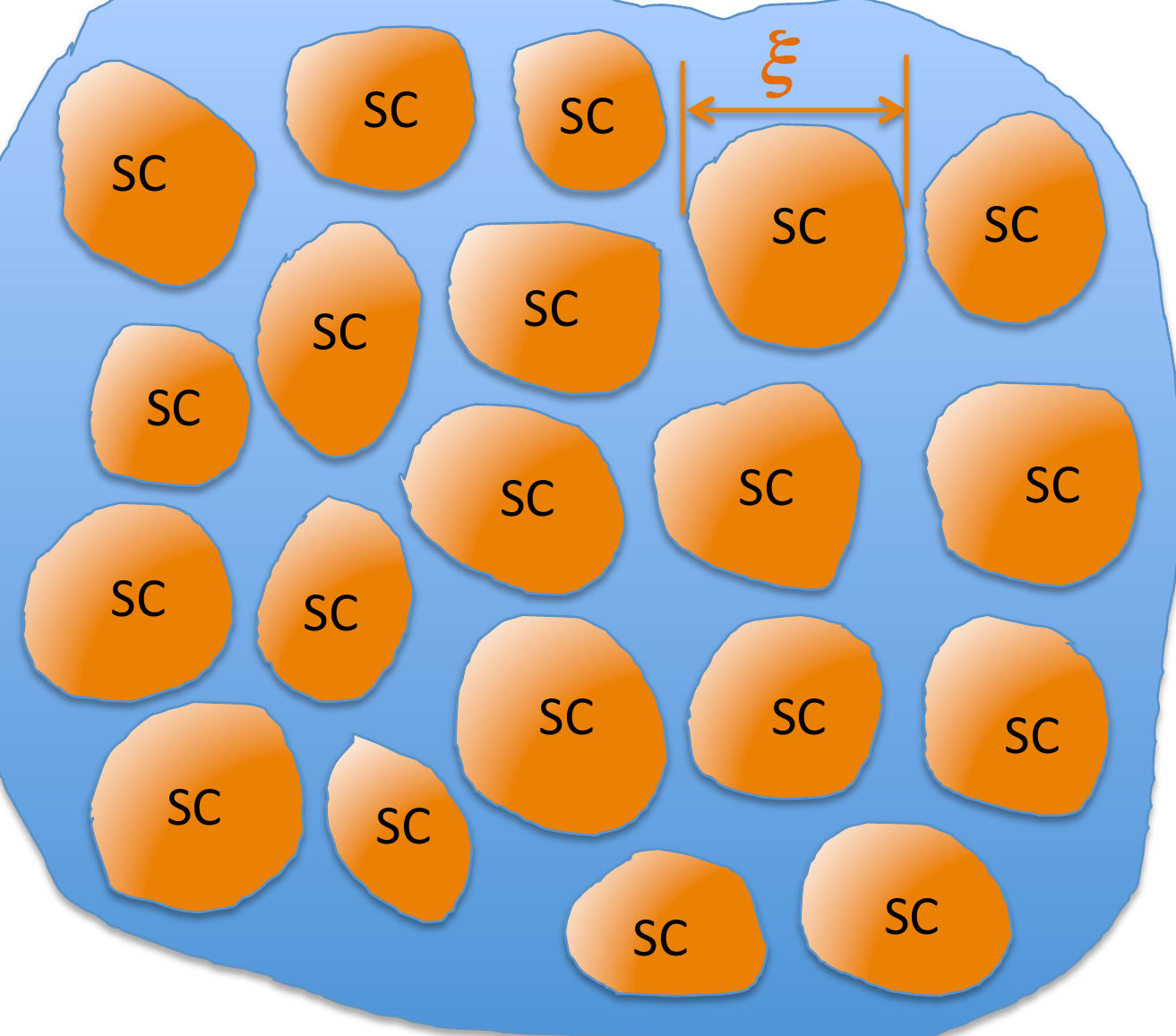}
\caption{\label{fig:Fig.5} {Self-induced electronic `granular' structure in the critical vicinity of the SIT.} At the first order direct superconductor-superinsulator transition ($\eta<1)$ the phase separation occurs and the system breaks into the superconducting droplets immersed into an insulating matrix and connected by weak Josephson links. The characteristic size of a droplet is $\gtrsim\xi$.}
\end{figure}
%%%%%%%%%%%%%%%%%%%%%%%%%%%%%%%%%%%%%%%%%%%%%%%%%%%%%%%%

Being an exemplary laboratory for theoretical and experimental study of the SIT and emerging phases in its critical vicinity, Josephson junction arrays offer a perfect model for strongly disordered superconducting films\,\cite{fazio,Tinkham}. Moreover, a striking quantitative agreement between the experimentally measured characteristics of the superinsulating state in disordered TiN films\,\cite{Baturina2007} and its description in terms of a regular JJA\,\cite{Fistul2008} suggests that the close connection between these two systems goes well beyond the similarity of coarse-grained JJA and disordered film. The idea of emergent electronic granularity, i.e. that even in structurally homogeneous films the electronic texture in the critical vicinity of the SIT is an array of superconducting puddles connected by weak links and immersed in an insulating matrix was pioneered in\,\cite{Ovadyahu1994}. Over two past decades, this hypothesis evolved into a paradigmatic attribute of the SIT, see\,\cite{Pratap2017} and references therein. It was conjectured\,\cite{Baturina2007,vinokur2008superinsulator} that it is this emergent granularity, see Fig.\,\ref{fig:Fig.5}, that serves as a material platform for the superinsulating state. Yet, while being generally accepted, the concept of emergent granularity is not thoroughly justified. Our gauge approach to the SIT enables us to put this concept on a firm field-theoretical foundation.

To see this, note that there is a finite width strip embracing the line of the direct SIT at $g=g_c=1$ and $\eta<1$, where Cooper pairs and vortex condensates coexist. Calculating from (\ref{topdef}) at $g>1$ and $\eta<1$ the energy $E$ of a
superconducting droplet of perimeter $L$ immersed into a superinsulating matrix
 of area $A$, we find (see SM)
\begin{equation}
E \propto A -{1\over 4\pi} \left( 1-{1\over g^2}\right) L^2 + \sigma L \,,
\label{energy}
\end{equation}
where the boundary contribution arises due to fluctuation-induced charge and vortex excitations within the opposite condensates, respectively, and $\sigma$ is a numerical coefficient. The energy has a global minimum $E_{\rm global} \propto A/g^2$ when the superconductor droplet fills the whole area $A$, but also a local minimum $E_{\rm local} \propto A$ at $L=0$. These two minima are separated by a maximum at $L \propto g^2/(g^2-1)$. If close enough to the SIT, this maximum spreads over a scale exceeding the radius of the droplet. Hence, it becomes energetically advantageous to fragment a superconducting droplet into smaller ones.
The fragmentation stops at the minimal dimension of order $\xi$ setting the scale of the self-induced granularity.

%\section*{Discussion}
There is a deep connection between the structure of the superinsulating state and field theory models of confinement\,\cite{confinement}. To illustrate this, we consider again the effective action for electromagnetic gauge fields. %While in the superconducting phase this amounts simply to a photon mass, equivalent to the London equations of superconductivity, 
In the superinsulating phase it is given by (see SM)
\begin{equation}
S^{\rm eff} \left ( A_{\mu}  \right) = {g\mu \eta \over 2\pi^2} \ \sum_{x, \mu} \big[ 1-{\rm cos} \left( 2e\ell^2 F_{\mu} \right) \big] \ ,
\label{pol}
\end{equation}
where $F_{\mu} = k_{\mu \nu} A_{\nu}$ is the dual field strength. 
This action coincides with Polyakov's compact quantum electrodynamics (QED) model\,\cite{polyakov}, in which a non-perturbative photon mass and linear confinement appear due to instanton effects. 
One can thus view the dipole Cooper pair-anti-Cooper pair neutral excitations in the superinsulator as neutral ``U(1) meson" excitations\,\cite{confinement}. The charge BKT transition of the SIT corresponds then to the deconfinement transition of the (2+1)-dimensional compact gauge theory\,\cite{yaffe}. 
This analogy has far reaching implications.  
In particular, depending on their spin quantum number, these neutral bound states of Cooper pairs could decay into, or mix with photons via the one-loop vacuum polarization diagram (see SM). In this case, applying an external electric field along the film would cause this to emit light and/or flicker. This effect calls for further investigation.

\subsection*{Acknowledgements}
We are delighted to thank M. Vasin and Ya. Kopelevich for illuminating discussions.
M.\,C.\,D. thanks CERN, where she completed this work, for kind hospitality. The work of V.\,M.\,V. was supported by the U.S. Department of Energy, Office of Science, Materials Sciences and Engineering Division.

%%%%%%%%%%%%%%%%%%%%%%%%%%%%%%%%%%%%%%%%%%%%%%%%%%%

\end{document}